\documentclass[longauth,letter, traditabstract]{aa} % for the long lists of affiliations

\usepackage{tabularx}
\usepackage{graphicx}
\usepackage{my_natbib}
\usepackage{txfonts}
\usepackage{longtable}
\usepackage{multirow}
\usepackage{lscape}

\newcommand{\hersc}{{\it Herschel}}

\newcommand{\spitz}{{\it  Spitzer}}

\newcommand{\iras}{{\it IRAS}}

\newcommand{\msun}{$M_\odot$}
\newcommand{\zsun}{$Z_\odot$}
\newcommand{\mic}{$\mu$m}

\renewcommand*{\figurename}{ Fig.}
\renewcommand*{\tablename}{Table}
\newlength{\pointwidth}
\def\revised{}

\begin{document}

  \title{\hersc\ photometric observations of the nearby low metallicity irregular galaxy NGC 6822 \thanks{\hersc\ is an ESA space observatory with science instruments provided by Principal Investigator consortia. It is open for proposals for observing time from the worldwide astronomical community.}}
  	      
\author{ 
	 M. Galametz\inst{1},
	 S. C. Madden\inst{1},
	 F. Galliano\inst{1},
	 S. Hony\inst{1},
	  M. Sauvage\inst{1},
	  M. Pohlen\inst{2},
	  G. J. Bendo\inst{3},
	  R. Auld\inst{2},
	  M. Baes\inst{4},
	  M. J. Barlow\inst{5},
	  J. J. Bock\inst{6},
	 A. Boselli\inst{7},
	  M. Bradford\inst{6},
	  V. Buat\inst{7},
	  N. Castro-Rodr{\'\i}guez\inst{8},
	  P. Chanial\inst{1},
	  S. Charlot\inst{9},
	  L. Ciesla\inst{7},
	  D. L. Clements\inst{3},
 	 A. Cooray\inst{10},
	  D. Cormier\inst{1},
	  L. Cortese\inst{2},
	  J. I. Davies\inst{2},
	  E. Dwek\inst{11},
	  S. A. Eales\inst{2},
	  D. Elbaz\inst{1},
	  W. K. Gear\inst{2},
           J. Glenn\inst{12},
	  H. L. Gomez\inst{2},
	  M. Griffin\inst{2},
	  K. G. Isaak\inst{13},
	  L. R. Levenson\inst{6},
	  N. Lu\inst{6},
	  B. O'Halloran\inst{3},
	  K. Okumura\inst{1},
	  S. Oliver\inst{14},
	  M. J. Page\inst{15},
	  P. Panuzzo\inst{1},
	  A. Papageorgiou\inst{2},
	  T. J. Parkin\inst{16},
	  I. P{\'e}rez-Fournon\inst{8},
	  N. Rangwala\inst{12},
	  E. E. Rigby\inst{17},
	  H. Roussel\inst{9},
	  A. Rykala\inst{2},
	  N. Sacchi\inst{18},
	  B. Schulz\inst{19},
	  M. R. P. Schirm\inst{16},
	  M. W. L. Smith\inst{2},
	  L. Spinoglio\inst{18},
	  J. A. Stevens\inst{20},
	  S. Sundar\inst{9},
	  M. Symeonidis\inst{15},
	  M. Trichas\inst{3},
	  M. Vaccari\inst{21},
	  L. Vigroux\inst{9},
	  C. D. Wilson\inst{16},
	  H. Wozniak\inst{22};
	  G. S. Wright\inst{23};
	  W. W. Zeilinger\inst{24}
          }

\institute{	
	CEA, Laboratoire AIM, Irfu/SAp, Orme des Merisiers, F-91191 Gif-sur-Yvette, France
		\email{ maud.galametz@cea.fr}
	\label{inst1}
	\and
	School of Physics $\&$ Astronomy, Cardiff University, Queens Buildings The Parade, Cardiff CF24 3AA, UK
          \label{inst2}
\and
	Astrophysics Group, Imperial College, Blackett Laboratory, Prince Consort Road, London SW7 2AZ, UK
	\label{inst3}
\and
	Sterrenkundig Observatorium, Universiteit Gent, Krijgslaan 281 S9, B-9000 Gent, Belgium
	 \label{inst4}
\and
 	Dept. of Physics $\&$ Astronomy, University College London, Gower Street, London WC1E 6BT, UK
	 \label{inst_5}
\and
	Jet Propulsion Laboratory, CA 91109; Dept. of Astronomy, California Institute of Technology, CA 91125;  Pasadena, USA
	 \label{inst_6}
\and	
	Laboratoire d'Astrophysique de Marseille, UMR6110 CNRS, 38 rue F. Joliot-Curie, F-13388 Marseille France
         \label{inst_7}
\and
	Instituto de Astrof{\'\i}sica de Canarias (IAC) and Dept. de Astrof{\'\i}sica, Universidad de La Laguna (ULL), La Laguna, Tenerife, Spain
	\label{inst_8}
\and
	Institut d'Astrophysique de Paris, UMR7095 CNRS, Univ. Pierre $\&$ Marie Curie, Boulevard Arago, F-75014 Paris, France
	\label{inst_9}
\and
	Dept. of Physics $\&$ Astronomy, University of California, Irvine, CA 92697, USA
	\label{inst_10}
\and
	Observational  Cosmology Lab, Code 665, NASA Goddard Space Flight  Center Greenbelt, MD 20771, USA
	\label{inst11}
\and
	Dept. of Astrophysical $\&$ Planetary Sciences, CASA CB-389, Univ. of Colorado, Boulder, CO 80309, USA
	\label{inst12}
\and
	ESA Astrophysics Missions Division, ESTEC, PO Box 299, 2200 AG Noordwijk, The Netherlands
	\label{inst_13} 
\and
	Astronomy Centre, Department of Physics  $\&$ Astronomy, Univ. of Sussex, UK
	\label{inst_14}
\and
	Mullard Space Science Laboratory, University College London, Holmbury St Mary, Dorking, Surrey RH5 6NT, UK
	\label{inst_15}
\and
	Dept. of Physics $\&$ Astronomy, McMaster University, Hamilton,  Ontario, L8S 4M1, Canada
	\label{inst_16}
 \and
	School of Physics $\&$ Astronomy, Univ. of Nottingham, University Park, Nottingham NG7 2RD, UK
	\label{inst_17}
\and
	Istituto di Fisica dello Spazio Interplanetario, INAF, Via del Fosso del Cavaliere 100, I-00133 Roma, Italy
	\label{inst_18}
\and
	Infrared Processing $\&$ Analysis Center, California Institute of Technology, Mail Code 100-22, 770 South Wilson Av, Pasadena, CA 91125, USA
	\label{inst_19}
\and
	Centre for Astrophysics Research, Univ. of Hertfordshire, College Lane, Herts AL10 9AB, UK
	\label{inst_20}
\and
	University of Padova, Department of Astronomy, Vicolo Osservatorio  
3, I-35122 Padova, Italy
	\label{inst_21}
\and
	Observatoire Astronomique de Strasbourg, UMR 7550 Univ. de Strasbourg - CNRS, 11, rue de l'Universit\'e, F-67000 Strasbourg
	\label{inst_22}
\and
	UK Astronomy Technology Center, Royal Observatory Edinburgh, Edinburgh, EH9 3HJ, UK 
	\label{inst23}
\and
	Institut für Astronomie, Universität Wien, Türkenschanzstr. 17,  A-1180 Wien, Austria
	\label{inst_24}
}

%===============================================================
% Abstract
%===============================================================

 \abstract
{
We present the first \hersc\ PACS and SPIRE images of the low-metallicity galaxy NGC6822 observed from 70 to 500 \mic\ and clearly resolve the H~{\sc ii} regions {\revised with PACS and SPIRE}. We find that the ratio 250/500 is dependent on the 24 \mic\ surface brightness in NGC6822, which would locally link the heating processes of the coldest phases {\revised of dust in the ISM to the star formation activity}. We {\revised model the SEDs} of some regions H~{\sc ii} regions and less active regions across the galaxy and find that the SEDs of H~{\sc ii} regions {\revised show} warmer ranges of dust temperatures. We derive {\revised very} high dust masses when graphite is used in our model to describe carbon dust. Using amorphous carbon, instead, requires less dust mass to account for submm emission due to its lower emissivity properties. This indicates that SED models including \hersc\ constraints {\revised may} require different dust properties than commonly used. The global G/D of NGC6822 is finally estimated to be 186, using amorphous carbon.}


     \authorrunning{M. Galametz et al}
     \titlerunning{NGC6822 revealed by the \hersc\ Space Observatory }

 \maketitle

    %--------------------- 3-color image ---------------------------------------------------------
\begin{figure}
   \centering
       \begin{tabular}{ c}
       
         \includegraphics[width=7.3cm ,height=6.3cm]{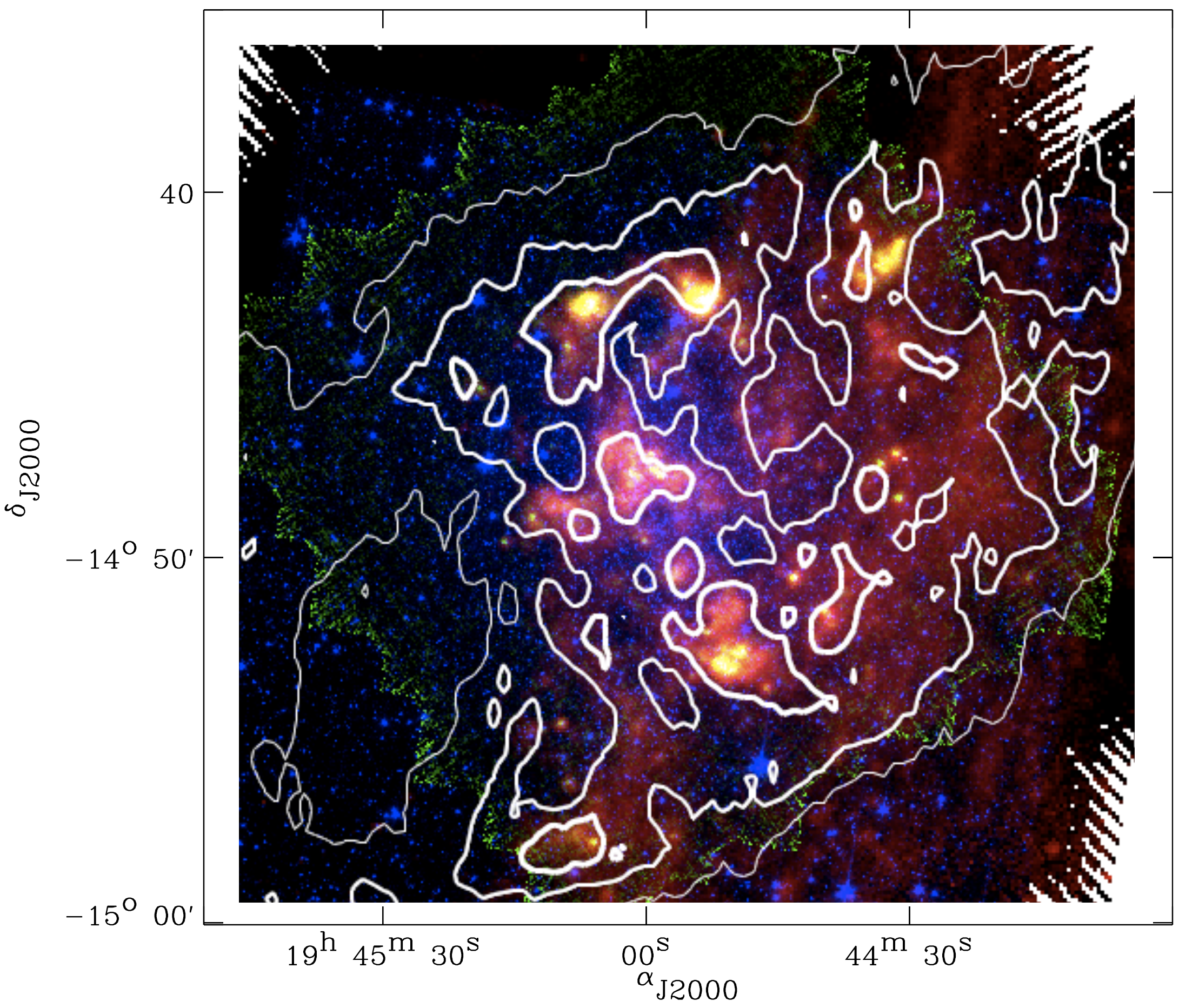} \\
        
         \end{tabular}
    \caption{NGC6822 \hersc/\spitz\ 3-color image. North is up, East is left. {\it Blue:} Stellar emission observed with \spitz/IRAC 3.6 \mic. {\it Yellow:} Warm dust emission with Herschel/PACS 100 \mic. {\it Red:} Cold dust emission with Herschel/SPIRE 250 \mic. H~{\sc i} contours at 3.1, 7.8 and 13 $\times$ 10$^{20}$ H.cm$^{-2}$ are overlaid \citep{deBlok2000}. The two bright compact IR knots in the North are Hubble X (east) and Hubble V (west).  }
    \label{3_color_image}
\end{figure}
%----------------------------------------------------------------------------------------------------------------------------

    %--------------------- PACS images ---------------------------------------------------------
\begin{figure}
   \centering
  \begin{tabular}{ m{3.6cm} m{4cm} }
         
         \includegraphics[width=4cm ,height=3.2cm]{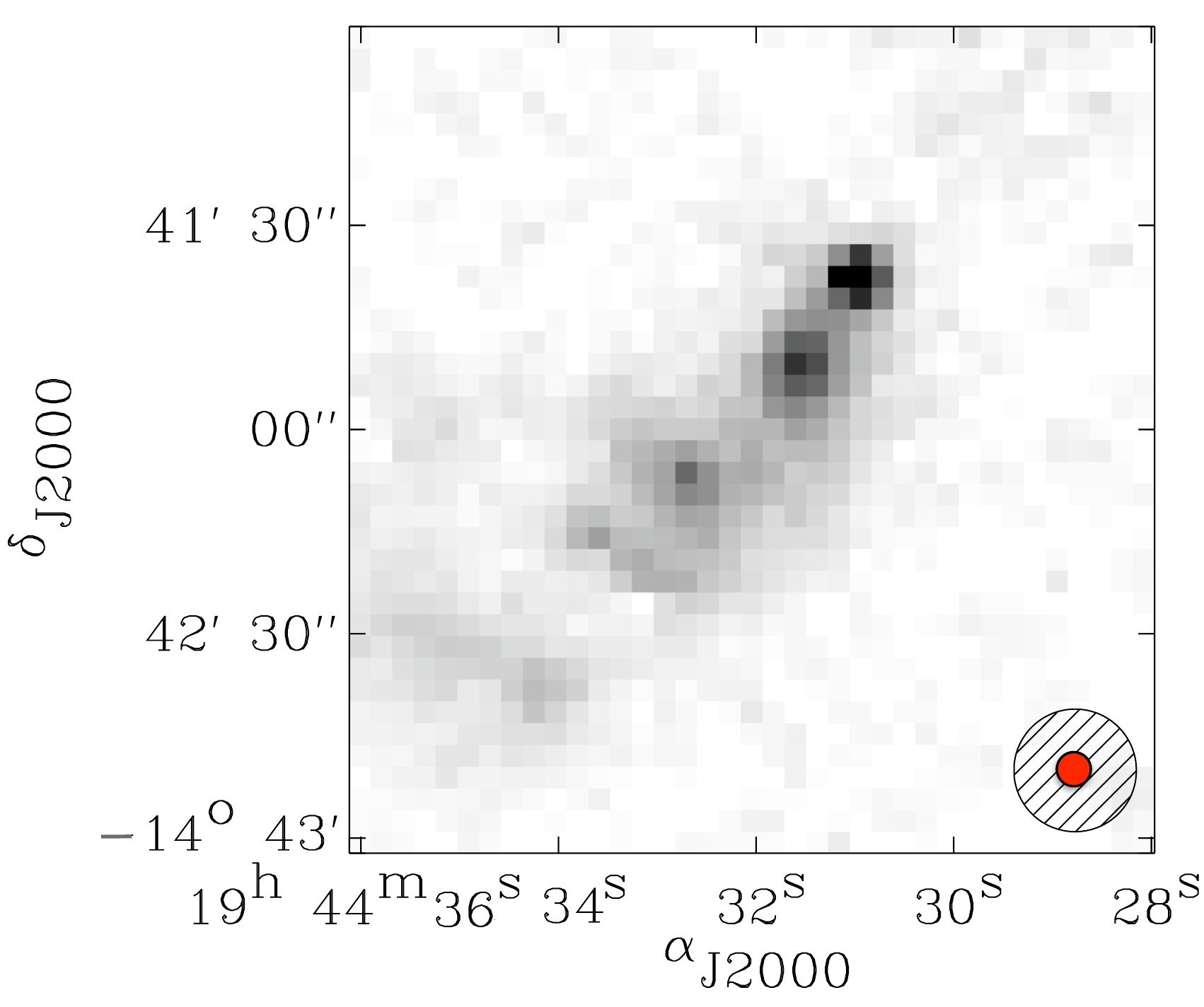}  &    							
         		\includegraphics[width=3.8cm ,height=3.2cm]{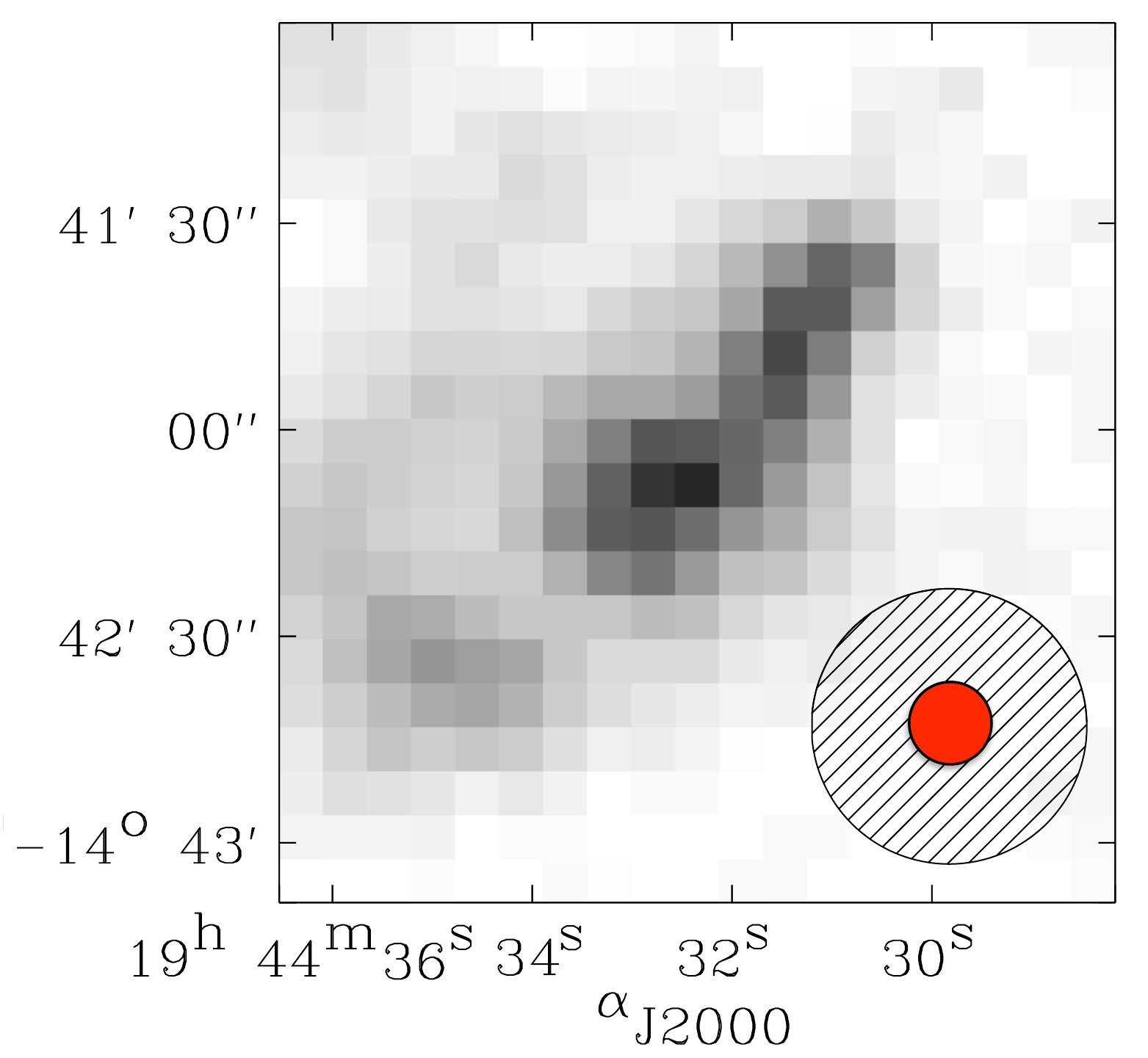}  \\
         \includegraphics[width=4cm ,height=3.2cm]{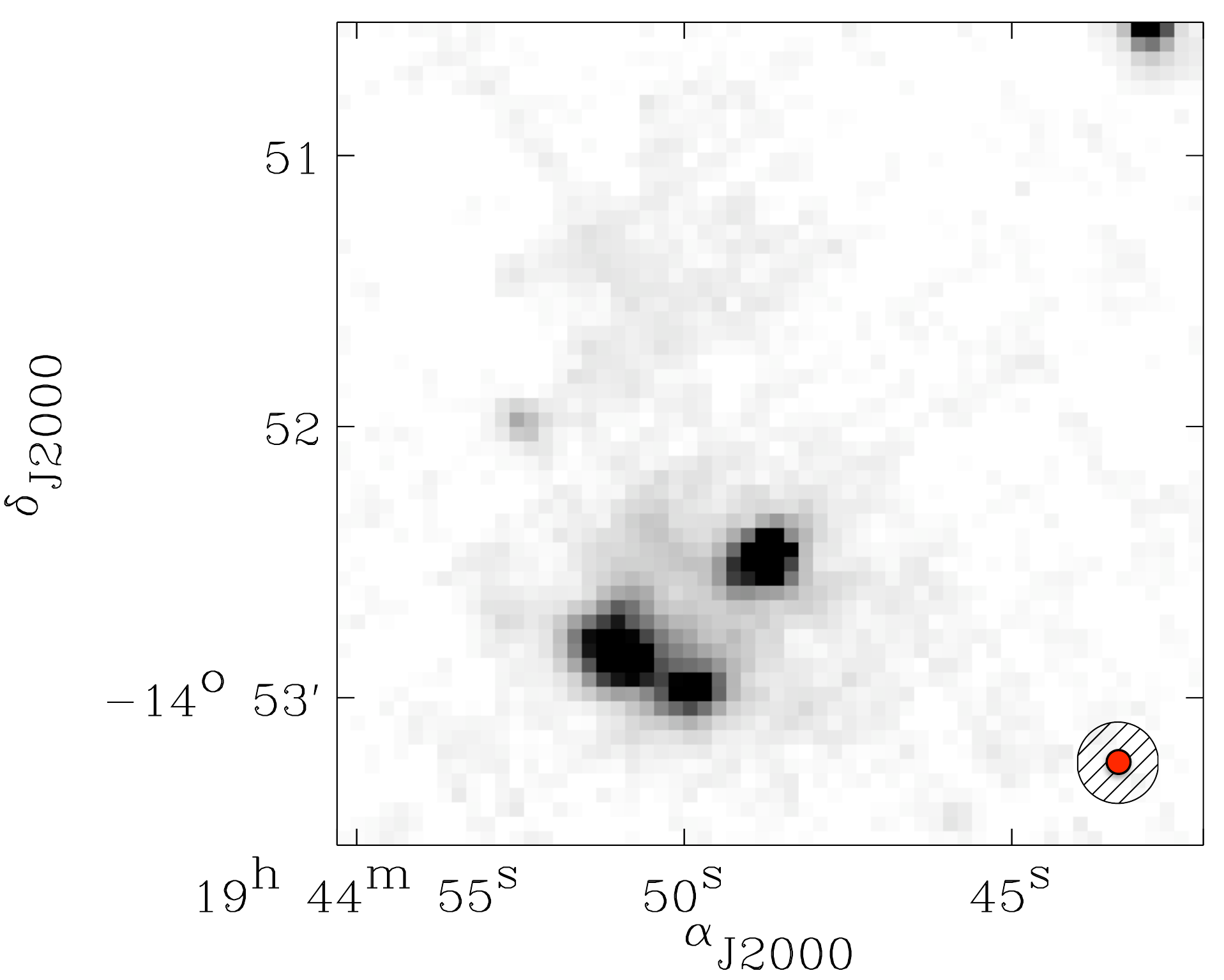}	&    							
        	 	\includegraphics[width=3.8cm ,height=3.2cm]{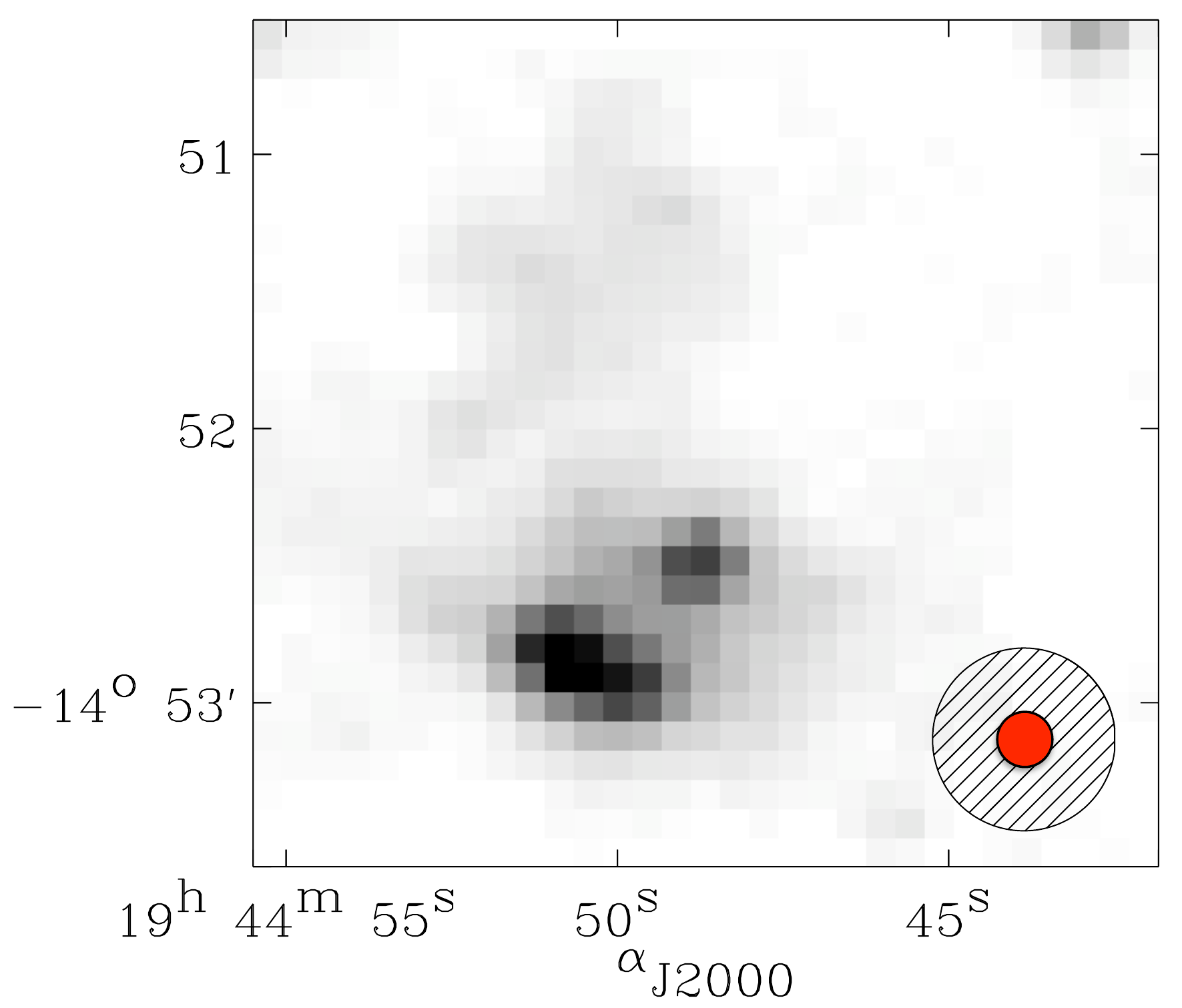}  \\
         \includegraphics[width=4cm ,height=3.2cm]{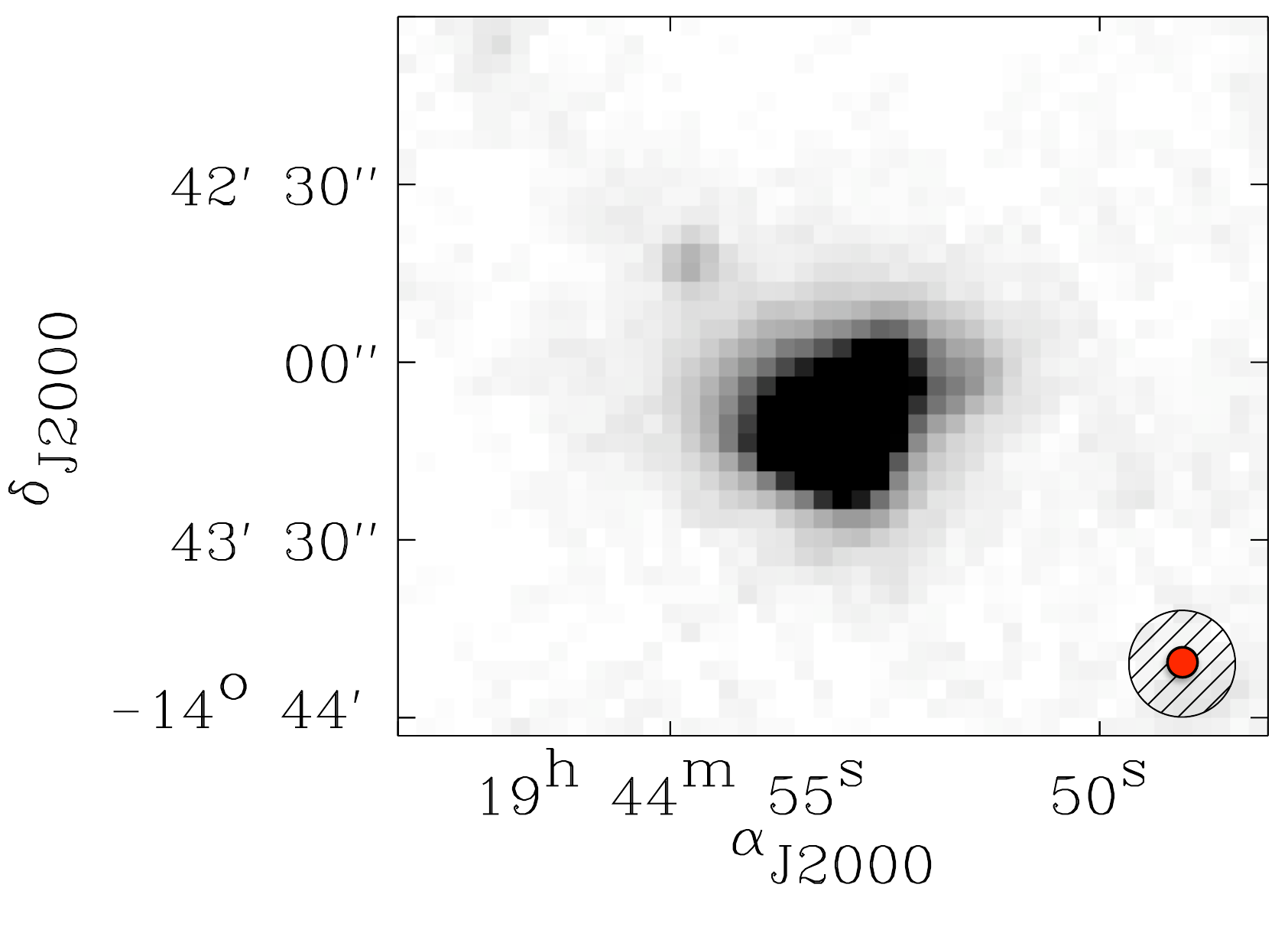}	&    							
        	 	\includegraphics[width=3.8cm ,height=3.2cm]{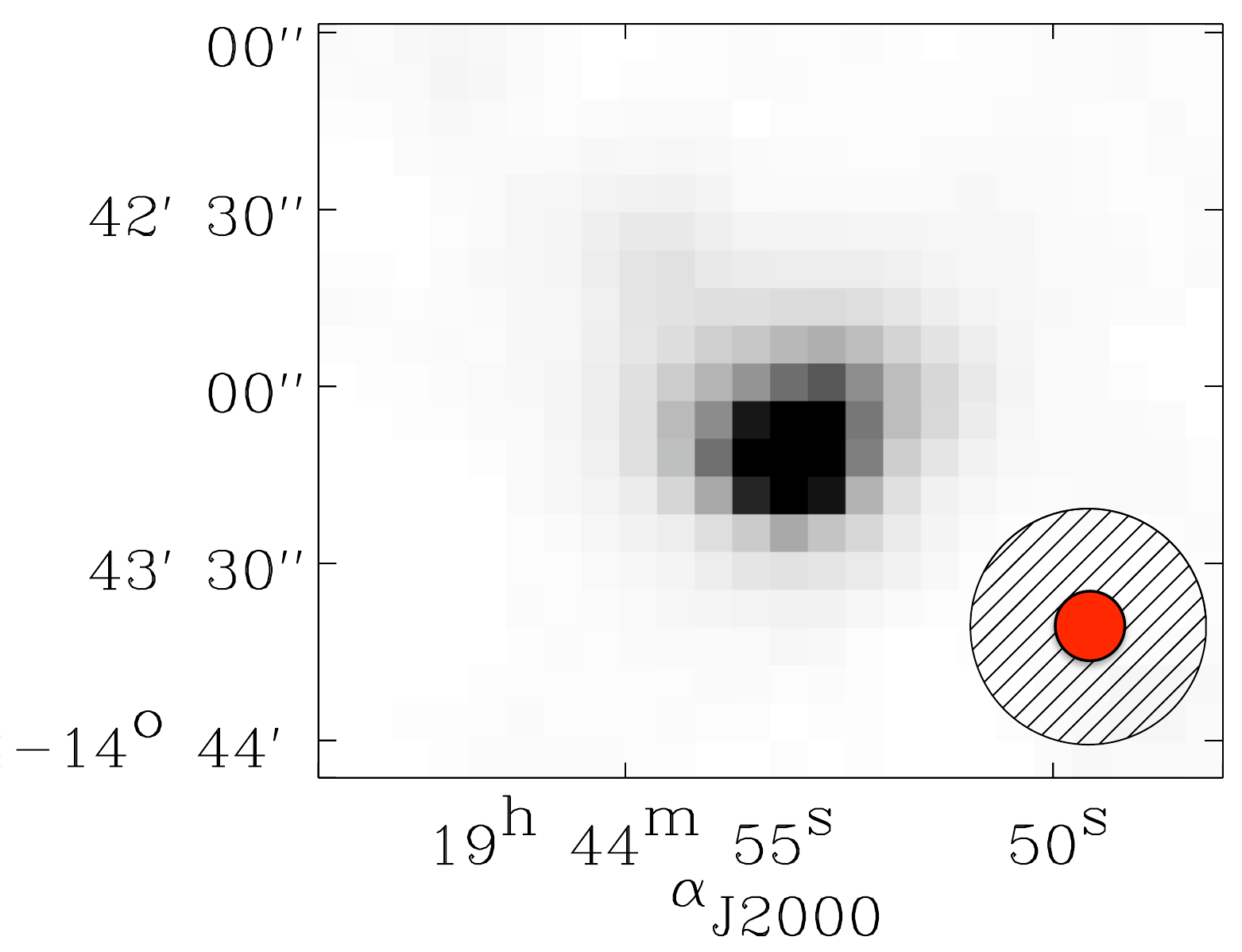}	\\
         \includegraphics[width=4cm ,height=3.2cm]{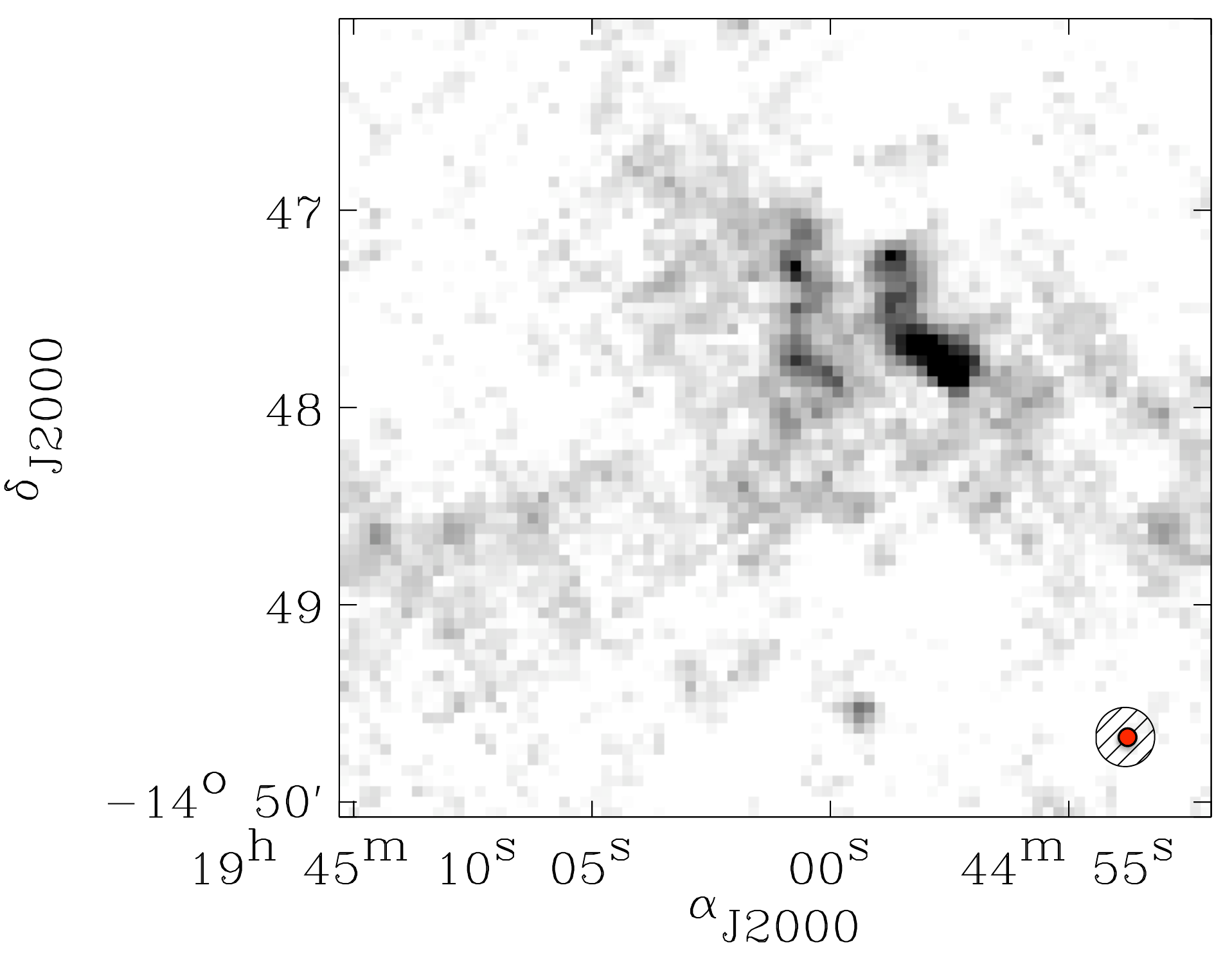}&    							
        	 	\includegraphics[width=3.8cm ,height=3.2cm]{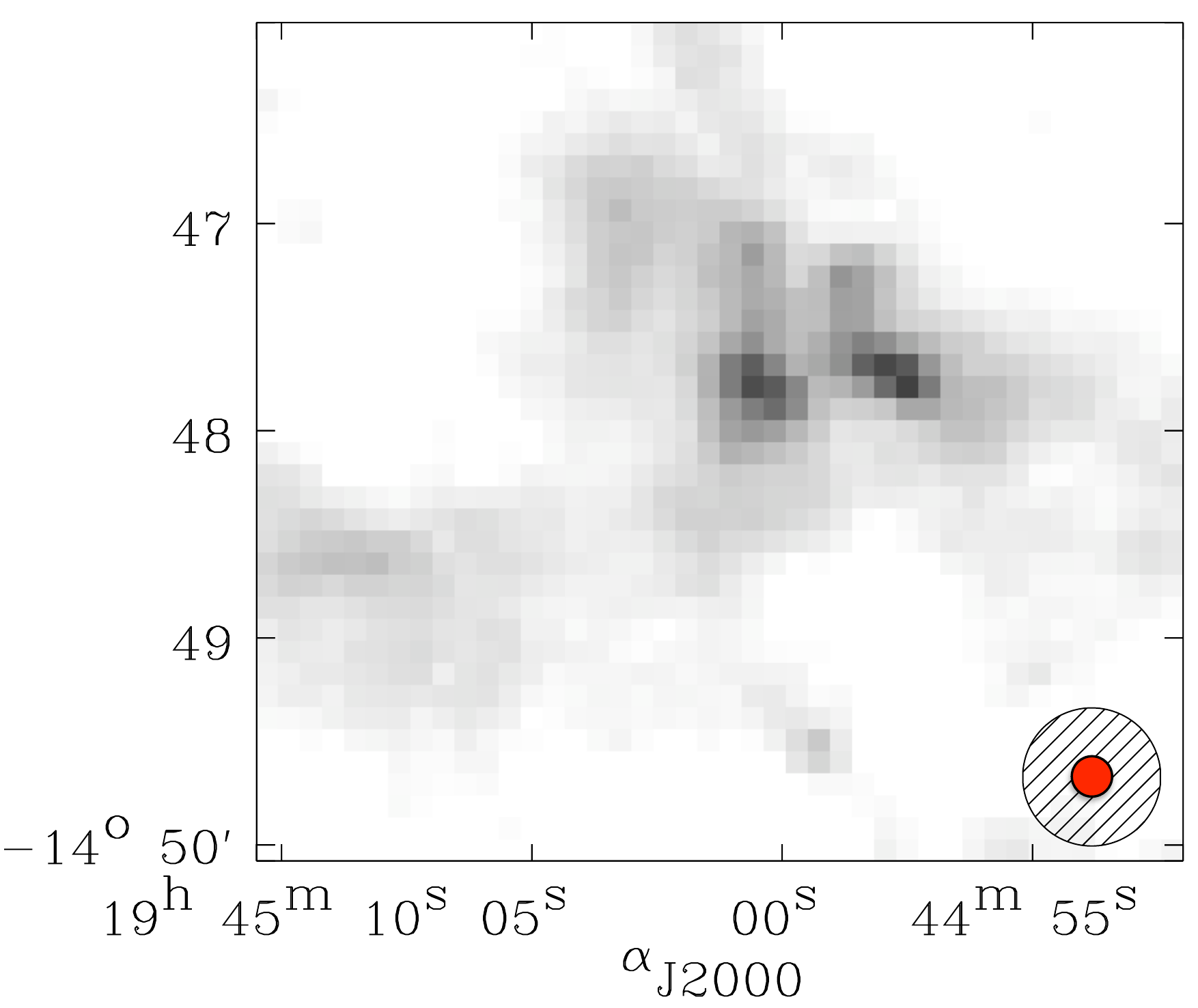} \\
         \includegraphics[width=4cm ,height=3.2cm]{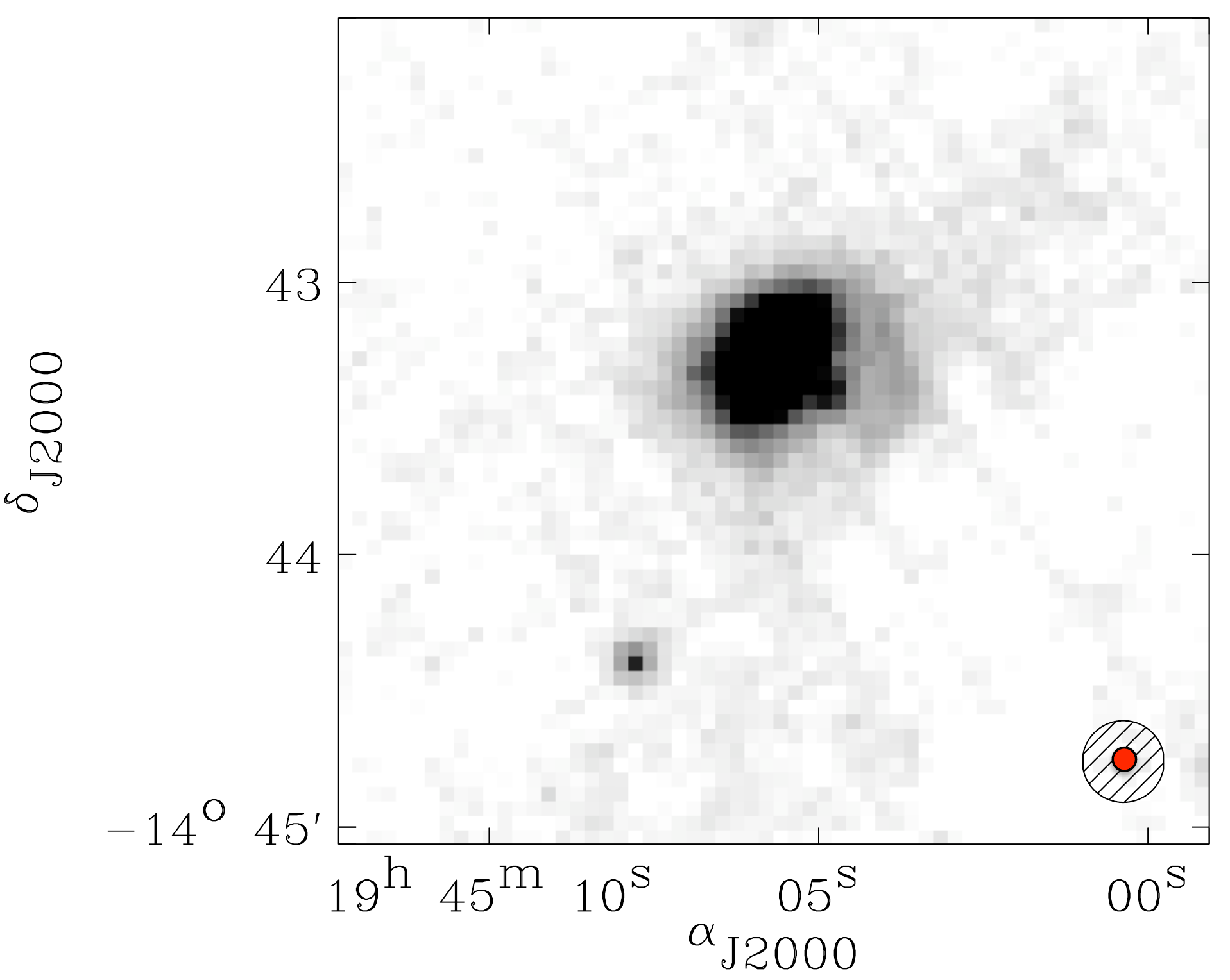} &   							
        	 	\includegraphics[width=3.8cm ,height=3.2cm]{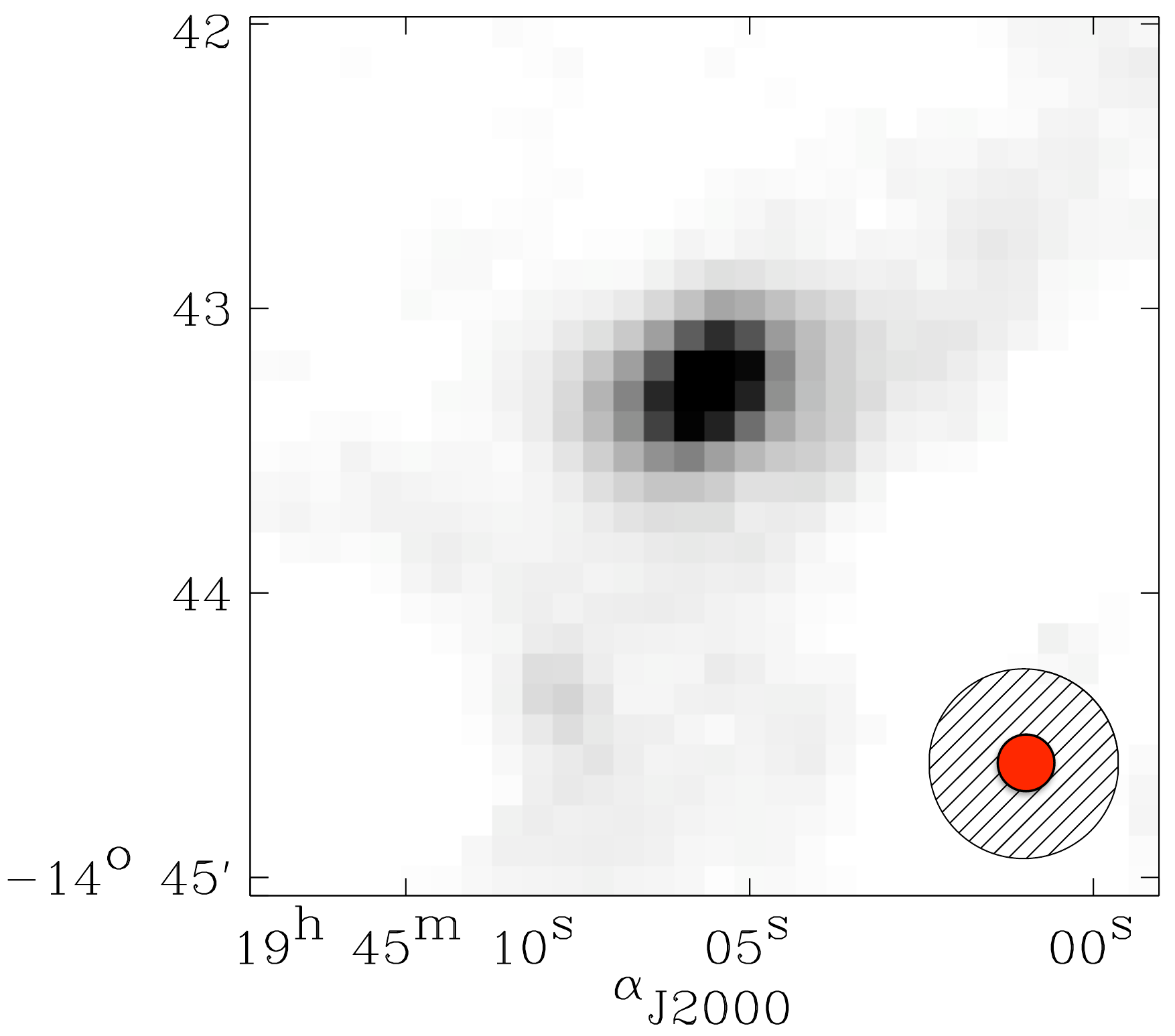}    \\
        
         \end{tabular}
    \caption{PACS observations of bright H~{\sc ii} regions of NGC6822. From top to bottom: Hubble I-III, Hubble IV, Hubble V, Hubble VI-VII and Hubble X (numbered 3, 7, 8, 11 and 14 in this paper - see Fig.3a), observed with PACS at 70 \mic\ (left) and 160 \mic\ (right). The FWHM  of \spitz/MIPS 70 and 160 \mic\ PSFs (black striped circles) and those of PACS 70 and 160 \mic\ (red circles) are overlaid for comparison.}
    \label{PACS_Images}
\end{figure}
%----------------------------------------------------------------------------------------------------------------------------

%------------------------------------------------------------------------------
%
%                     Introduction
%
%------------------------------------------------------------------------------

\section{Introduction}

The absorption of stellar radiation and its reemission by dust at infrared (IR) wavelengths is a fundamental process controlling the heating and cooling of the interstellar medium (ISM). The \iras, {\it ISO} and \spitz\ IR space telescopes launched the studies of the physics and chemistry of dust and gas, revealing their roles in the matter cycle and thermal balance in galaxies. The mid-infrared (MIR) to far-infrared (FIR) wavelength windows provide the necessary observational constraints on the spectral energy distribution (SED) modelling of galaxies from which properties of the polycyclic aromatic hydrocarbons (PAHs) and the warm ($>$30K) and hot dust can be determined.
Now, the \hersc\ Space Observatory \citep{Pilbratt2010} is probing the submillimeter regime, where the coldest phases ($<$30K) of dust can be revealed. 

Dwarf galaxies of the Local Group are {\revised nearby} laboratories to spatially study the life cycle of the different dust components and the metal enrichment of low-metallicity ISM. NGC6822 is our closest \citep[490 kpc;][]{Mateo1998} metal-poor galaxy neighbour ($\sim$ 30$\%$ \zsun) beyond the Magellanic Clouds and possesses isolated star forming (SF) regions at different evolutionary stages. The galaxy is a perfect candidate to study the feedback of the star formation on the low-metallicity ISM by analysing the spatial variations of its SEDs. NGC6822 also possesses an intriguing rotating H~{\sc i} disk of 1.34 $\times$ 10$^{8}$ \msun, that extends far beyond the optical disk \citep{Mateo1998} and has one of the largest HI holes ever observed in a dwarf galaxy \citep{deBlok2000}. 

NGC6822 was observed in 2009 October as part of the Science Demonstration observations for the Dwarf Galaxy Survey (PI :  S. Madden), with the instruments PACS and SPIRE at 70, 100, 160 and 250, 350 and 500 \mic\ respectively. At SPIRE 500 \mic\ (36"), we can resolve ISM structures of $\sim$ 85pc, spatially sufficient to accurately probe the distribution of dust temperature throughout the galaxy, especially dust in its coldest phases.

%------------------------------------------------------------------------------
%
%                     Herschel Maps
%
%------------------------------------------------------------------------------

\section{Observations and data reduction}

PACS \citep{Poglitsch2010} observations were performed in cross scan map mode at 70, 100 and 160 \mic. The observations cover a region of 18' $\times$ 18' around the starforming complexes of the galaxy also mapped with \spitz\ \citep{Cannon_NGC6822_2006}. Data reduction was carried out using a modified \hersc\ Interactive Processing Environment (HIPE) 3.0 pipeline, starting from the level 0 data and produced maps with pixel sizes of 3.2, 3.2 and 6.4" at 70, 100 and 160 \mic\ with PSF FWHM values of 5.2, 7.7 and 12" respectively.  
HIPE is used to suppress the bad pixels and those affected by saturation and convert the signal to Jy.pixel$^{-1}$. We perfom flatfield correction and apply astrometry to the data cube. We then apply a multiresolution median transform (MMT) deglitching correction and a second order deglitching process to the data. We perform polynomial fits on half scan legs to subtract the baselines. The galaxy was masked in the data cube during this step to avoid an overestimation of the signal level on the source. The median baseline subtraction step {\revised suppresses most of the bolometer temperature drifts}. We generate the final maps using the madmap method of HIPE. The absolute flux calibration uncertainties are estimated to be $\pm$10$\%$.

SPIRE \citep{Griffin2010} observations, at 250, 350 and 500 \mic, cover a region of 26' $\times$ 26'. Data have been reprocessed from the level 1 cube following the steps described in \citet{Pohlen2010} and \citet{Bendo2010}. The overall absolute calibration accuracy is estimated to be $\pm$ 15$\%$ \citep{Swinyard2010}. The pipeline produces maps with a pixel size of 6", 10" and 14" at 250, 350 and 500 \mic\ with PSF FWHM values of 18", 25" and 36" respectively. The {\revised SPIRE} ICC has released preliminary multiplying factors {\revised to correct for extended sources}: 1.02, 1.05 and 0.94 at 250, 350 and 500 \mic\ \footnote{As advised by the ICC for SD papers, see http:// herschel.esac.esa.int/SDP$\_$wkshops/presentations/IR/3$\_$Griffin$\_$SPIRE
$\_$SDP2009.pdf}.

%---------------------SPIRE images ---------------------------------------------------------
\begin{figure*}
   \centering
       \begin{tabular}{ p{0cm} m{5.1cm} p{0cm} m{5.1cm} p{0cm} m{7cm}}
        a)  &\includegraphics[width=4.9cm ,height=4.4cm]{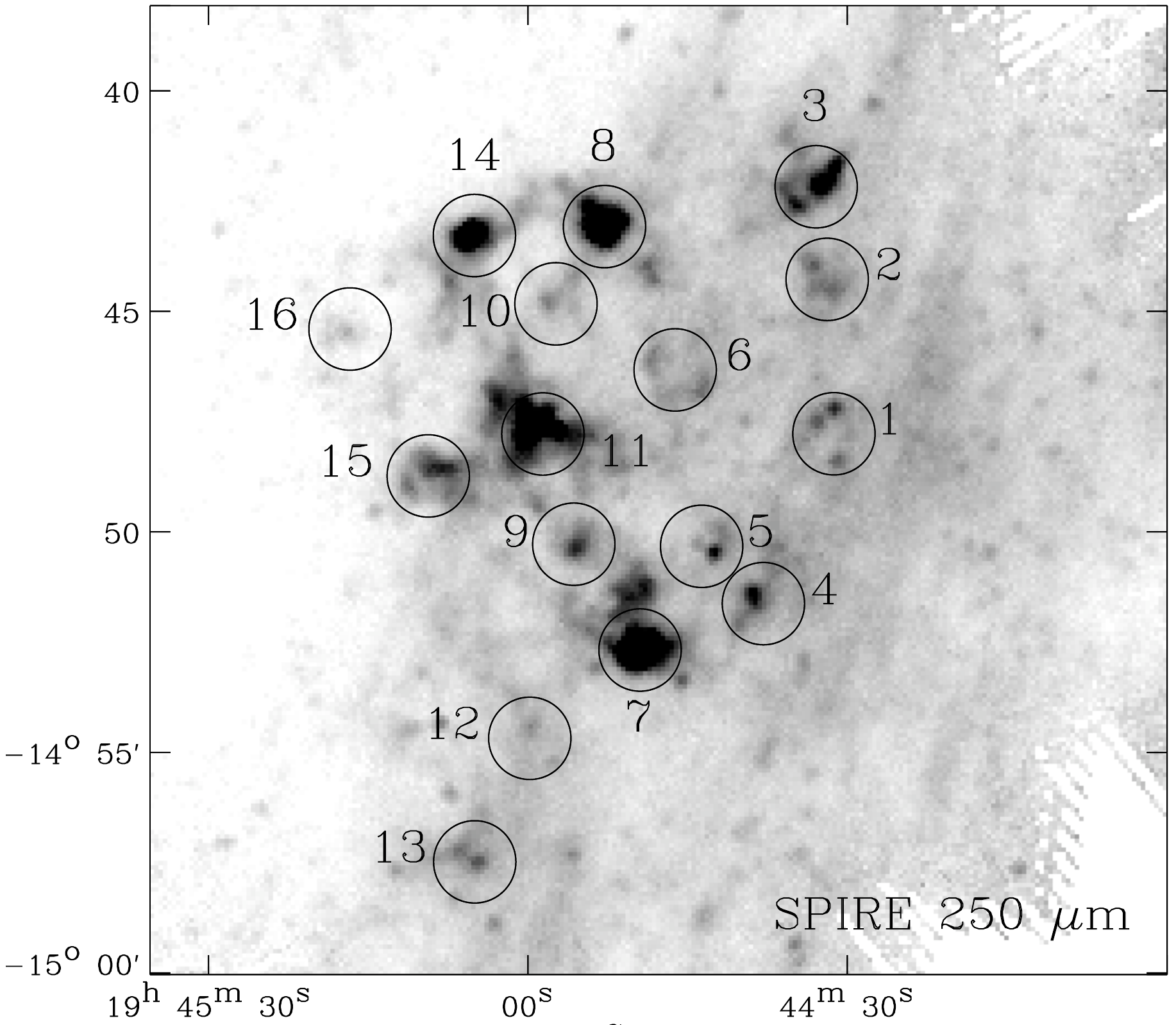} &
         b) & \includegraphics[width=4.9cm ,height=4.4cm]{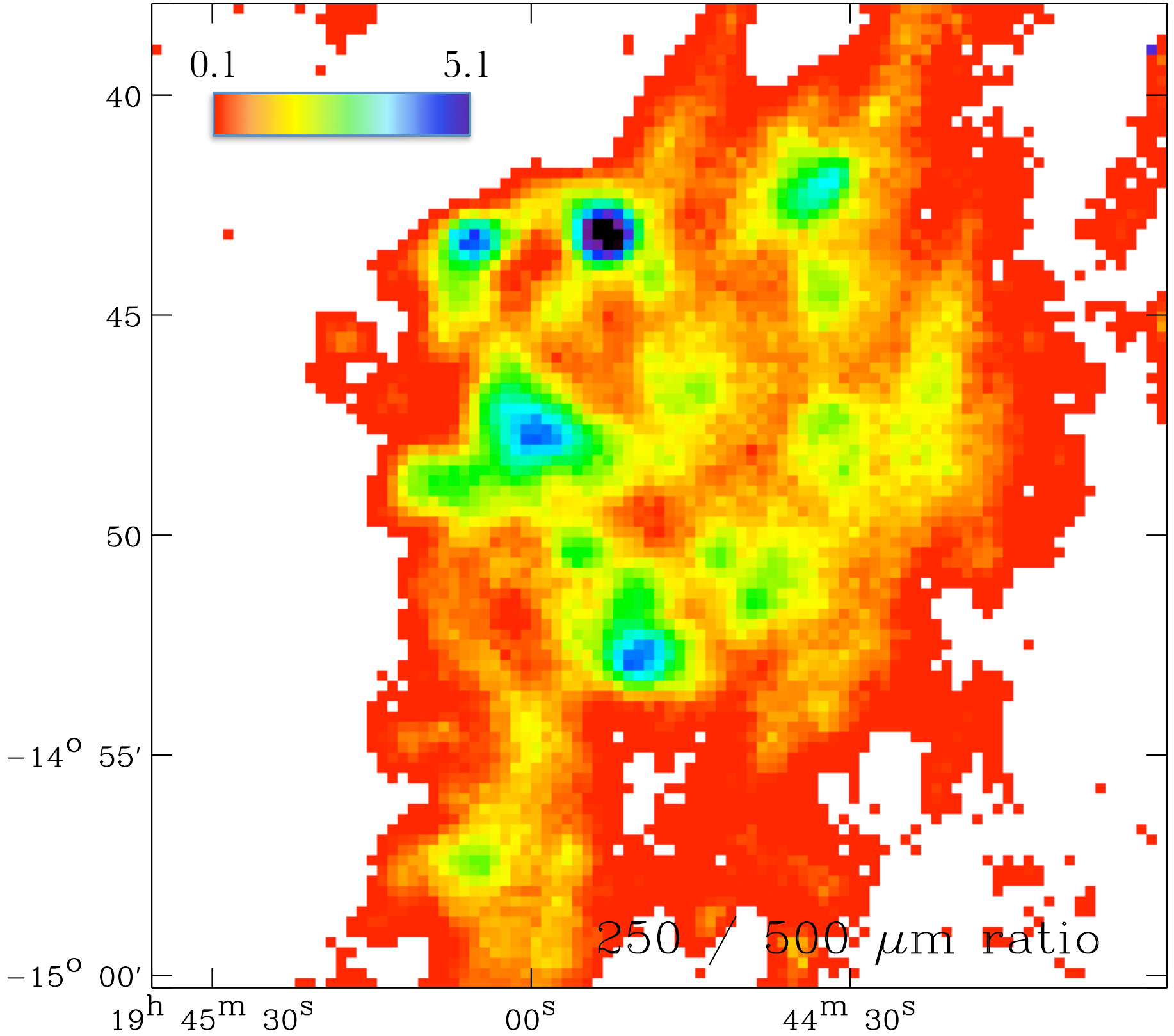} &
          c) & \includegraphics[width=5.5cm ,height=4.4cm]{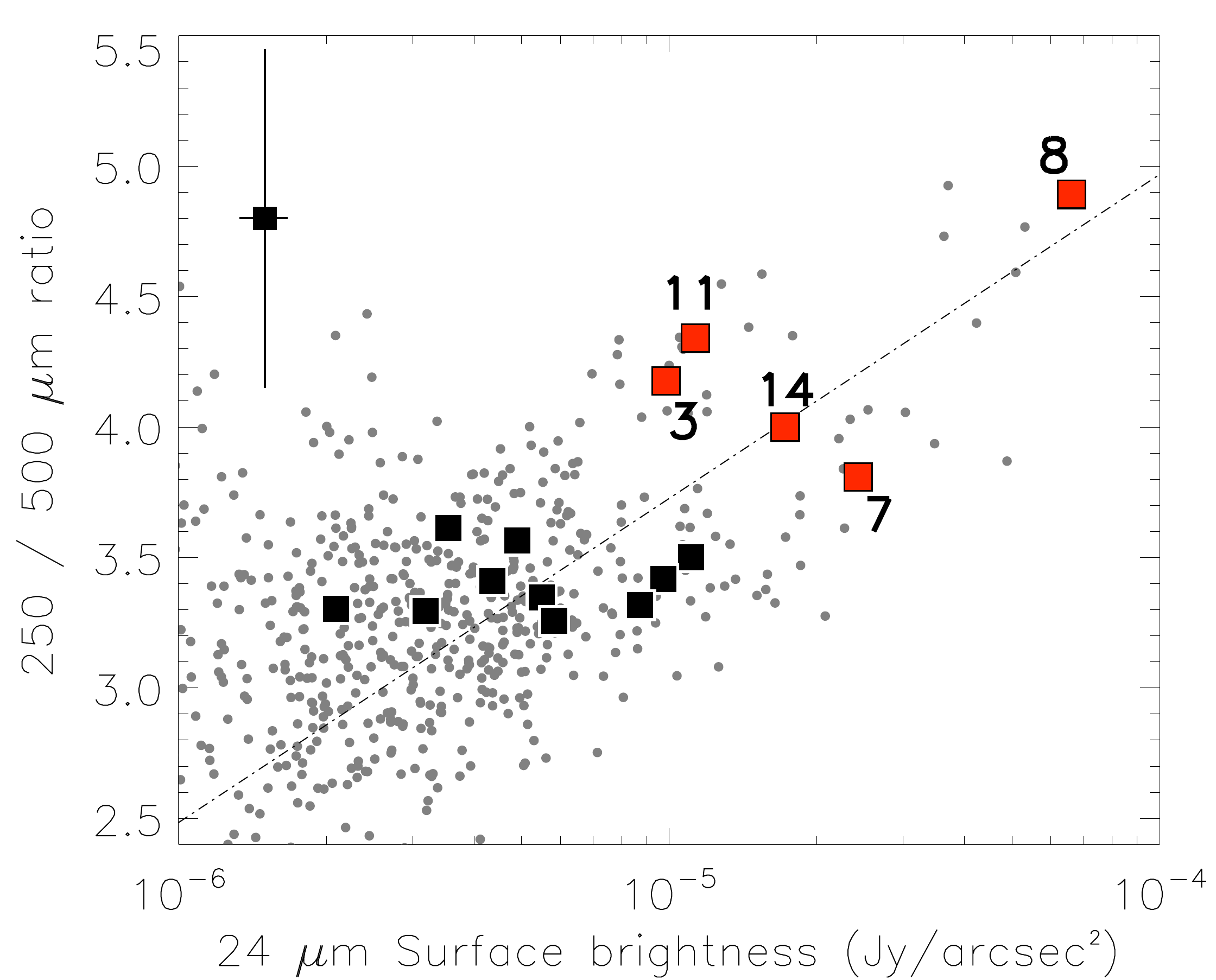} \\
         
         \end{tabular}
    \caption{a) NGC6822 observed at 250 \mic. Circles indicate our photometric apertures (57" radius). b) 250 / 500 \mic\ flux density ratio map of NGC6822. c) F$_{\nu}$(250) / F$_{\nu}$(500) ratio as a function of the 24 \mic\ surface brightness. Ratios are estimated in the apertures of Fig.3a (squares) {\revised and pixel-by-pixel (grey circles). In this second case, pixels are chosen to have the size of the FWHM of SPIRE 500 \mic}. Red squares indicate the brightest H~{\sc ii} regions. Since the uncertainties in the SPIRE fluxes are preliminary, we give an indication of the error bars on the upper left corner. }
    \label{SPIRE}
\end{figure*}
%----------------------------------------------------------------------------------------------------------------------------

%------------------------------------------------------------------------------
%
%       			 Analysis
%
%------------------------------------------------------------------------------

\section {Analysis}

\subsection {\hersc\ maps}

Figure~\ref{3_color_image} is a color composition (size of the SPIRE maps) showing how stars, dust and H~{\sc i} gas (in contours) are distributed in the galaxy. IRAC 3.6 \mic\ {\revised band} (blue) mostly traces the stellar emission while PACS 100 \mic\ (green) traces the warm dust. SPIRE 250 \mic\ (red) traces the {\revised cold dust phases}. We find that the bright H~{\sc ii} regions are resolved at all \hersc\ wavelengths. The brightest IR/submm SF regions all coincide with H~{\sc i} peaks. {\revised While H~{\sc i} extends 30' to the NW and SE of the galaxy, the big H~{\sc i} hole is devoid of dust emission.} The star formation history of NGC6822 began 12-15 Gyr ago and has been quiescent until about 0.6-1 Gyr ago \citep{Gallart1996, Wyder2001}. \citet{Bianchi2001} found very young stellar populations ($<$10Myr) in the star-forming regions Hubble V and X. These regions are also the brightest structures of the FIR and submm emission (see the two brightest knots in the North of NGC6822 in Fig.~\ref{3_color_image}). 

PACS 70 and 160 \mic\ observations of the 5 brightest star-forming regions of NGC6822 (Hubble I-III, IV,  V, VI-VII and X) are shown in Fig.~\ref{PACS_Images}. These regions are respectively numbered 3, 7, 8, 11 and 14 in Fig.~\ref{SPIRE}a. The increase in spatial resolution from MIPS \citep{Cannon_NGC6822_2006} to PACS (see the respective FWHM PSFs in Fig.~\ref{PACS_Images}) now enables us to nicely separate the different substructures of the H~{\sc ii} regions Hubble I-III (3) and Hubble VI-VII (11) {\revised (see online material for complete images)} but also shows faint emission across the galaxy that was detected but not resolved with MIPS. To compare MIPS and PACS fluxes at 70 and 160 \mic\ for these H~{\sc ii} regions, we convolve the images to the lowest (FWHM MIPS 160 \mic: 40") resolution \citep{Bendo2010} and extract flux densities using the function {\it aper} of IDL in apertures of 57" radius ($\sim$135 pc). We find that, for the bright H~{\sc ii} regions of NGC6822, MIPS and PACS flux densities compare within $\pm$25$\%$ at 70 and 160 \mic\ {\revised but 33$\%$ for Hubble V at 70 \mic.} 
The SPIRE maps also resolve the structures detected with PACS. We find that the diffuse emission of NGC6822 is strongly affected by non homogeneous Galactic cirrus emission due to the low Galactic latitude of NGC6822 (low level diffuse emission in Fig.~\ref{SPIRE}a). 
{\revised We model the cirrus emission in each map as a plane after first masking the emission that we associate with the galaxy. We then remove the modeled cirrus emission from each SPIRE map for the analysis. }
The cirrus contamination is estimated to be less than 10$\%$ of the flux densities of the star-forming regions at 250 and 350 \mic\ . It is less than 15$\%$ and 25 $\%$ on the less active regions respectively at 250 \mic\ and 350 \mic. The contamination is finally estimated to be $\sim$ 25 $\%$ on the bright regions at 500 \mic\ but can be as high as $\sim$ 50 $\%$ of the emission in {\revised the lowest level of the diffuse ISM. Our quantitative results concern relatively bright regions. Thus,} taking into account the various uncertainties, we attribute an overall conservative estimate of $\sim$ 30$\%$ for all SPIRE bands. We note that our results are strongly dependant on the SPIRE fluxes, more precisely on our treatment of cirrus emission that could have been underestimated if some of this emission is more significant {\revised along some} lines of sight.

\subsection{SPIRE band ratios}

We convolve the 250 \mic\ map to the resolution of SPIRE 500 \mic\ (36") and build a 250 / 500 flux density ratio map (Fig.~\ref{SPIRE}b) to study the evolution of the submm regime of the SEDs. The ratio {\revised peaks} in the H~{\sc ii} regions (Hubble V showing the highest ratio). This ratio map highlights the evolution of the dust temperature distribution across the galaxy: warmer toward the H~{\sc ii} regions and decreasing in the diffuse extended dust component between the H~{\sc ii} regions. To understand the processes contributing to the heating of dust, we examine how the submm part of the SED evolves with star formation. We select individual regions and estimate the SPIRE 250 and 500 \mic\ flux densities of these regions in apertures of 57" radius {\revised (Fig.~\ref{SPIRE}a), corresponding to regions studied in} \citet{Cannon_NGC6822_2006}. In Fig.~\ref{SPIRE}c, we plot the 250 / 500 flux density ratio of the selected regions (squares) as a function of their 24 \mic\ surface brightness. {\revised The same 250/500 ratios performed on a pixel-by-pixel \footnote {The pixel size of the maps was chosen to equal the FWHM of SPIRE 500 \mic\ (36").} basis throughout the whole map are overlaid (grey circles).} The 24 \mic\ flux is commonly used as a tracer of star formation \citep[e.g.][]{Calzetti2007}. The 250/500 ratios seem to correlate with the 24 \mic\ surface brightness across NGC6822, which could imply that the cold dust temperature distribution varies with the star formation activity of the region, {\revised with higher temperature dust present} where star formation activity dominates \citep{Boselli2010}. \citet{Bendo2010} find that the SPIRE band ratios in M81depend on radius, and that the old stellar population of the bulge and disk could be the primary source for the dust emission seen by SPIRE. Their submm ratios do not show any strong correlation with {\revised 24 \mic} surface brightness. This effect suggests that the dust heating processes {\revised in low metallicity starbursts may differ from normal dusty spirals}. The ISM of dwarfs is, indeed, less opaque than spirals, with dense SF regions usually influencing the whole galaxy. These galaxies also {\revised tend to have a preponderance of younger stellar population and less evolved stars, and thus very different star formation histories.}

%  

%---------------------SED ---------------------------------------------------------
\begin{figure}
   \centering
       \begin{tabular}{ c}
         \includegraphics[width=8.1cm ,height=7.3cm]{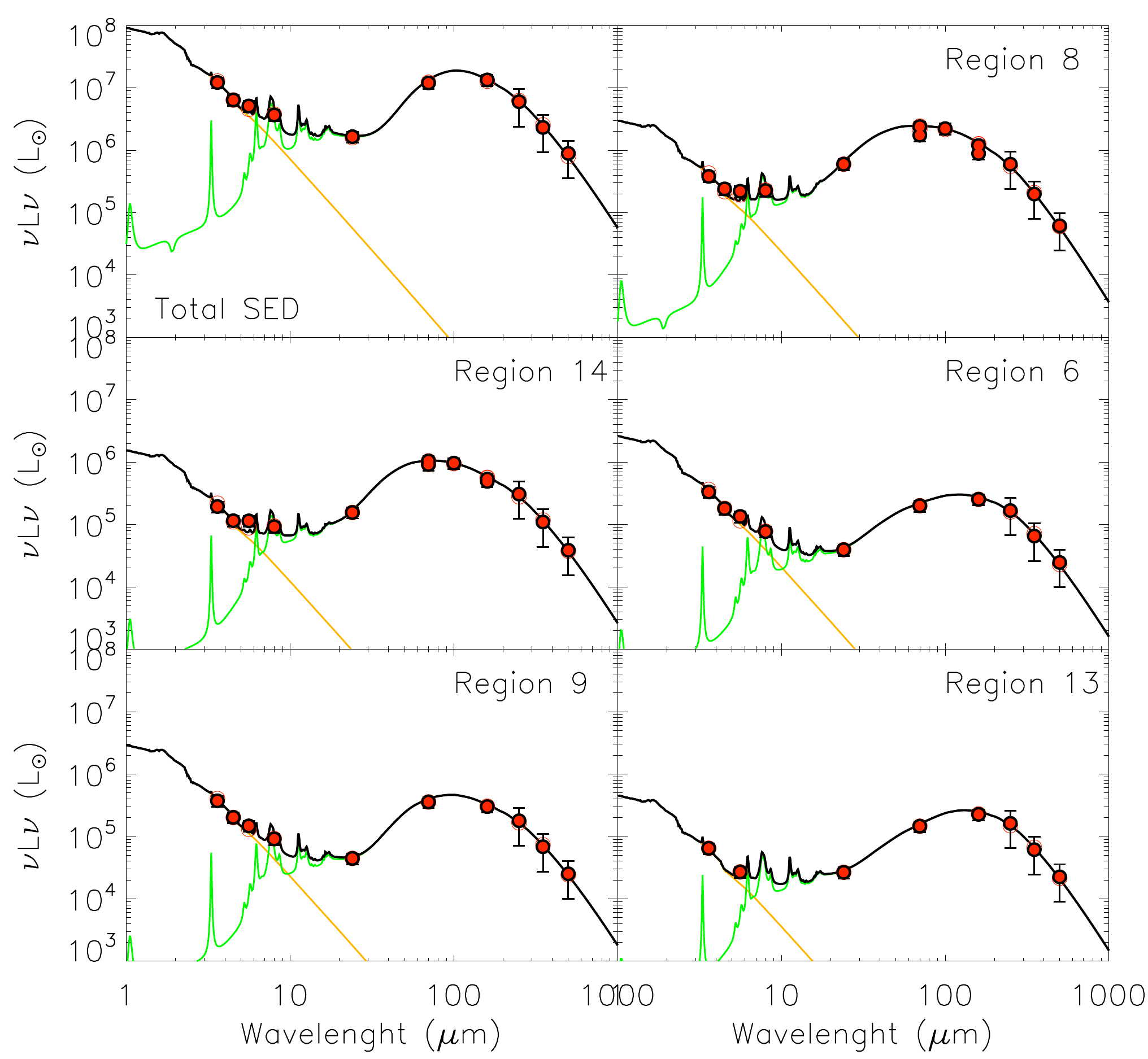} \\     
         \end{tabular}
    \caption{Total SED of NGC6822 and SEDs of individual regions among which bright H~{\sc ii} regions (Region 8 and 14) and less active regions (Regions 6, 9 and 13). Our SED models are plotted in black. Observational constraints are overlaid in red circles. The orange and green lines distinguish the stellar and the dust contribution. The 30$\%$ uncertainties are conservatively estimated for the SPIRE bands. }
    \label{SEDs}
\end{figure}
%----------------------------------------------------------------------------------------------------------------------------

%----------------------------Dust / Gas Table------------------------------------------------------------------------------------------------
\begin{table}
\caption{General properties}
\tiny
\label{Properties}
 \centering
 \begin{tabular}{c@{  \  \ }c@{\  \  \  \ }c@{\  \  \  \ }c@{\  \  \  \ }c@{\  \  \  \ }c}
\hline
\hline
Id             &    M$_{HI}$		   & M$_{dust~Gr}$ $^a$			 & M$_{HI}$ / M$_{dust~Gr}$       & M$_{dust~AC}$ 	$^a$		 & M$_{HI}$/ M$_{dust~AC}$		 \\
	       &	[10$^5$ M$_{\sun}$] & [10$^3$ M$_{\sun}$]	&		  & [10$^3$ M$_{\sun}$]  & \\
\hline

8   & 7.6 	&	$22.2^{+7}_{-6}$ \rule[-6pt]{0pt}{4pt}		& 34  & $7.8^{+3}_{-3}$ 		& 97	\\
14 & 9.1	&	$23.0^{+8}_{-7}$ \rule[-6pt]{0pt}{4pt}		& 39	& $10.2^{+4}_{-4}$ 		& 89\\
\hline
6  & 5.8 	& 	$14.1^{+5}_{-4}$ \rule[-6pt]{0pt}{4pt}		& 41 & $5.3^{+2}_{-2}$ 		& 109\\
9   & 7.7	&	$22.1^{+5}_{-5}$ \rule[-6pt]{0pt}{4pt}		& 34 & $8.5^{+3}_{-4}$ 		& 91\\ 
13 & 7.8 	&	$12.9^{+5}_{-3}$ \rule[-6pt]{0pt}{4pt}		& 60  & $4.7^{+3}_{-3}$ 		&166 \\
\hline
Total  & 500 $^b$ & $612^{+105}_{-182}$ \rule[-6pt]{0pt}{4pt}    & 80 & $269^{+143}_{-145}$  & 186	\\
\hline
\end{tabular}
\begin{list}{}{}
\item[$^a$] Dust mass derived using graphite (Gr) or amorphous carbon (AC).
\item[$^b$] H~{\sc i} mass corresponding to the region mapped with \hersc.
\end{list}
 \end{table} 
 %----------------------------------------------------------------------------------------------------------------------------

\subsection{SEDs}

To study the variations of the local SEDs, we select two H~{\sc ii} regions (Hubble V and X, numbered 8 and 14 in Fig.~\ref{SPIRE}a) and three less active regions (Reg. 6, 9 and 13). {\revised We suspect the median baseline filtering of the PACS reduction to remove} some of the diffuse emission. PACS fluxes are thus not used in the modelling of the total SED and less active regions. \spitz\ observations \citep[SINGS 5$^{th}$ release; ][]{Kennicutt2003} complete the coverage. We convolve the images to MIPS 160 \mic\ resolution (40") and estimate the flux densities within 57" radius apertures. 

We use a realistic SED model {\revised which follows the approach of the \citet{Dale2001} and \citet{Draine_Li_2007} models, and use the dust} composition and size distribution of \citet{Zubko2004}. The interstellar radiation field is assumed to have the shape of the Galactic diffuse ISM of \citet{Mathis1983}. The dust mass exposed to a given heating intensity U is given by: dM$_{dust}$(U)$\propto$U$^{-\alpha}$dU with U$_{min}$$<$U$<$U$_{max}$ \citep{Dale2001}. $\alpha$ parametrises the contribution of the different local SEDs exposed to U. Details on the modelling can be found in \citet{Galametz2009}. \citet{Serra_Dias_Cano2008} study carbon dust in shock waves and warn about {\revised using} graphites in dust models. To study how this choice affects our dust masses, we test both graphites and amorphous carbons \citep{Rouleau_Martin_1991} to describe the interstellar carbon dust.
Figure~\ref{SEDs} presents the global SED of NGC6822 along with the individual SEDs of the 5 selected regions obtained with amorphous carbons. No submm excess seems to be detected in NGC6822, contrary to other dwarf galaxies observed with \hersc\ \citep{OHalloran2010, Grossi2010}. We find that the SEDs of H~{\sc ii} regions {\revised have} warmer dust temperatures than less active regions. The total SED of NGC6822 does not show a high 24/70 ratio, indicating that it may not be dominated by the IR emission of bright H~{\sc ii} regions. The dust masses derived using graphites are 2.2 to 2.8 times higher than those using amorphous carbons (Table~\ref{Properties}), due to their lower emissivity at submm wavelengh.

To estimate the gas-to-dust mass ratios (G/D), we derive the H~{\sc i} mass of our regions from the integrated map of \citet{deBlok2000}. {\revised We also estimate the H~{\sc i} mass corresponding to the region mapped with \hersc\ to be $\sim$ 5 $\times$ 10$^7$ \msun}. \citet{Gratier2010} found that H$_2$ derived from CO observations should not represent more than 10$^7$ \msun. {\revised Faint emission lines from warm H$_2$ are observed in Hubble V \citep{Hunter_Kaufman_2007}}. \citet{Cannon_NGC6822_2006} note that the major H~{\sc ii} regions correspond to the strongest H$\alpha$ sources of the galaxy. From their H$\alpha$ fluxes, we derive an H$\alpha$ mass inferior to 10$^{5}$ \msun\ in the H~{\sc ii} regions \citep[assuming T=10$^4$K and N$_e$=100]{Storey1995}. H~{\sc i} thus dominates the gas mass in NGC6822. 
The \citet{Galliano_Dwek_Chanial_2008} models predicts G/D of $\sim$500 for galaxies presenting the metallicity of NGC6822. We find low G/D (Table~\ref{Properties}) for the individual regions compared to what can be expected from chemical evolution models, especially when graphite grains are used in the modelling. Amorphous carbon {\revised results in} a flatter submm slope compared to graphite and thus require{\revised s} less mass to produce the same emission. These results are consistent with those found by \citet{Meixner2010} in the Large Magellanic Cloud.
The total dust mass of the central region mapped with \hersc\ is 2.7 $\times$ 10$^5$ \msun\ using amorphous carbon dust, leading to a total G/D of 186.

%------------------------------------------------------------------------------
%
%                		 Conclusions
%
%------------------------------------------------------------------------------

\section{Conclusions}

We present \hersc\ images of NGC6822 which resolve ISM structures up to 500 \mic. We find that the 250/500 ratio (tracing the cold dust temperature range) may be dependent on the 24 \mic\ surface brightness and thus {\revised trace the} SF activity. We model individual SEDs across NGC6822 and show that the SED shape is evolving from H~{\sc ii} regions to less active regions, with H~{\sc ii} regions having a {\revised warmer dust temperature} range. We derive very high dust masses using graphite to describe carbon dust and find that the use of amorphous carbon decreases the dust masses, indicating that SED models including \hersc\ constraints require different dust properties, namely more emissive grains.

% {\revised Using amorphous carbon dust, we estimate the G/D of the region mapped by \hersc\ to be $\sim$186. }

%===============================================================
% Acknowledgements
%===============================================================

\begin{acknowledgements}
We thank the referee for his comments that help to improve the quality of this paper. We also thank Erwin de Blok for the integrated H~{\sc i} map of NGC6822. 
PACS has been developed by MPE (Germany); UVIE (Austria); KU Leuven, CSL, IMEC (Belgium); CEA, LAM (France); MPIA (Germany); INAF-IFSI/OAA/OAP/OAT, LENS, SISSA (Italy); IAC (Spain). This development has been supported by BMVIT (Austria), ESA-PRODEX (Belgium), CEA/CNES (France), DLR (Germany), ASI/INAF (Italy), and CICYT/MCYT (Spain). SPIRE has been developed by Cardiff University (UK); Univ. Lethbridge (Canada); NAOC (China); CEA, LAM (France); IFSI, Univ. Padua (Italy); IAC (Spain); SNSB (Sweden); Imperial College London, RAL, UCL-MSSL, UKATC, Univ. Sussex (UK) and Caltech, JPL, NHSC, Univ. Colorado (USA). This development has been supported by CSA (Canada); NAOC (China); CEA, CNES, CNRS (France); ASI (Italy); MCINN (Spain); Stockholm Observatory (Sweden); STFC (UK); and NASA (USA).
\end{acknowledgements}

%===============================================================
% Bibliography
%===============================================================

\bibliographystyle{aa}

\tiny{\bibliography{/Users/mgalamet/Documents/PhD/Papers/mybiblio.bib}}

%---------------------Online material ---------------------------------------------------------

\newpage

\begin{figure*}
   \centering
       \begin{tabular}{c}
         \includegraphics[width=18cm ,height=19cm]{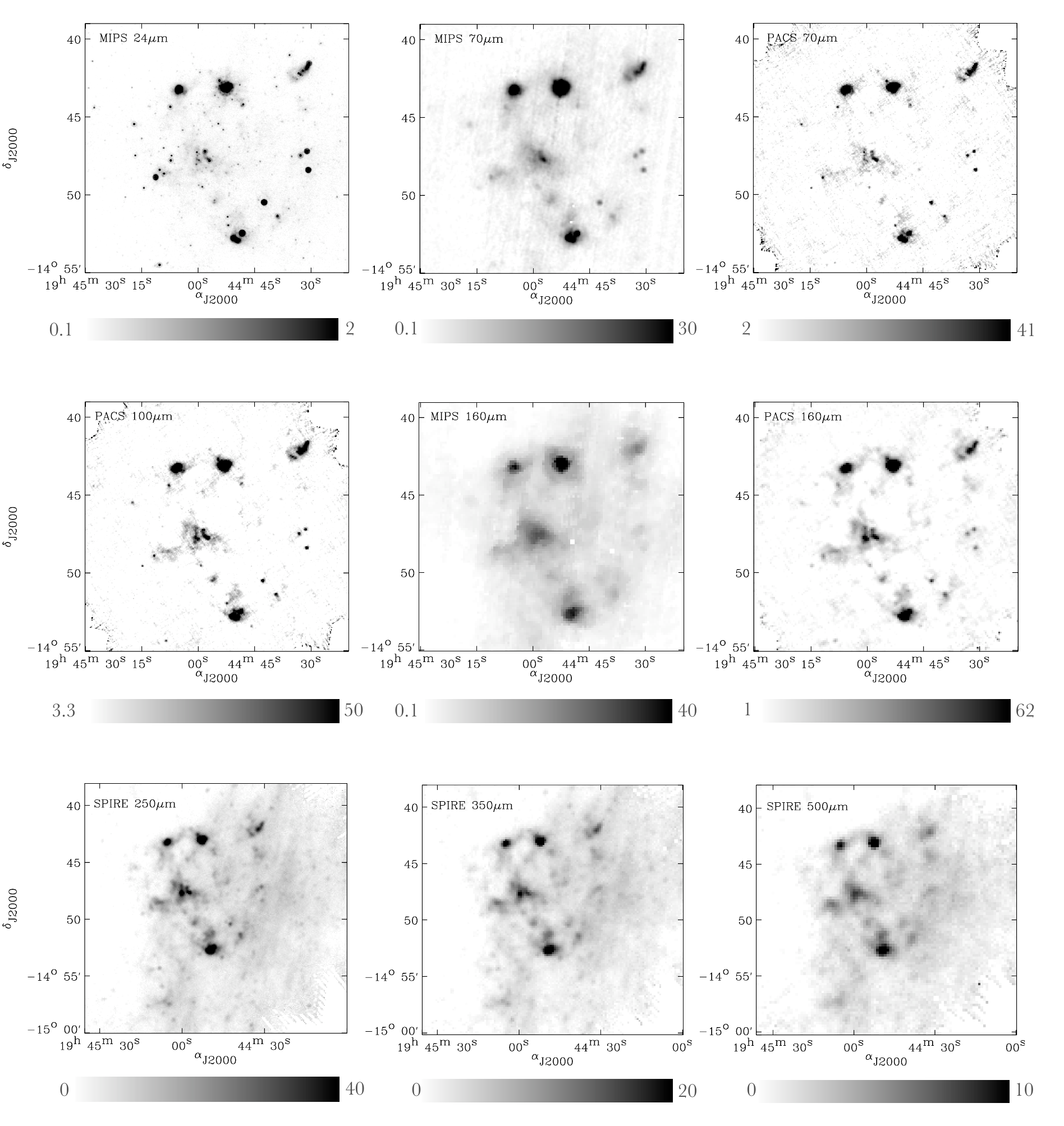} \\     
         \end{tabular}
    \caption{NGC6822 observed by \spitz/MIPS and \hersc/PACS and SPIRE. Fluxes are in MJy/sr.}
    \label{Full_images}
\end{figure*}
%----------------------------------------------------------------------------------------------------------------------------

\end{document}